%% file: main.tex
\documentclass[manuscript,screen]{acmart}
\AtBeginDocument{%
  \providecommand\BibTeX{{%
    \normalfont B\kern-0.5em{\scshape i\kern-0.25em b}\kern-0.8em\TeX}}}

\setcopyright{acmcopyright}
\copyrightyear{2024}
\acmYear{2024}
\acmDOI{XXXXXXX.XXXXXXX}

\acmConference[TBD 'XX]{TBD 'XX: Not published in the ACM}{September 6, 2024}{TBD}
\acmBooktitle{TBD 'XX: Not published in the ACM, September 6, 2024, TBD}
\acmISBN{978-1-4503-XXXX-X/18/06}

\usepackage{enumitem}
\usepackage[normalem]{ulem} 
\usepackage{xstring} 

\newif\ifShowChanges
\ShowChangesfalse

\relax

\ifShowChanges
  \definecolor{darkcyan}{HTML}{0090D8}
  \newcommand{\add}[1]{\textcolor{darkcyan}{\textbf{#1}}} 
  \newcommand{\strike}[1]{\sout{#1}} 
\else
  \newcommand{\add}[1]{{#1}} 
  \newcommand{\strike}[1]{\StrGobbleRight{}{1}} 
\fi

\begin{document}

\title[Data Dump to Digestible Chunks]{From Data Dump to Digestible Chunks: Automated Segmentation and Summarization of Provenance Logs for Communication}


\author{Jeremy E. Block}
\email{j.block@ufl.edu}
\orcid{0000-0003-1626-9074}
\author{Donald Honeycutt}
\email{dhoneycutt@ufl.edu}
\author{Brett Benda}
\email{brett.benda@ufl.edu}
\orcid{0000-0002-1825-6392}
\author{Benjamin Rheault}
\email{brheault@ufl.edu}
\orcid{0009-0009-8530-6861}
\author{Eric D. Ragan}
\email{eragan@ufl.edu}
\orcid{0000-0002-7192-3457}
\affiliation{%
  \institution{Computer \& Information Science \& Engineering\\
  University of Florida}
  \streetaddress{432 Newell Drive}
  \city{Gainesville}
  \state{Florida}
  \country{USA}
  \postcode{32611}
}

\renewcommand{\shortauthors}{Block, et al.}

\begin{abstract}
Communicating one's sensemaking during a complex analysis session to explain thought processes is hard, yet most intelligence occurs in collaborative settings.
Team members require a deeper understanding of the work being completed by their peers and subordinates, but little research has fully articulated best practices for analytic provenance consumers.
This work proposes an automatic summarization technique that separates an analysis session and summarizes interaction provenance as textual blurbs to allow for meta-analysis of work done.
Focusing on the domain of intelligence analysis, we demonstrate our segmentation technique using five datasets, including both publicly available and classified interaction logs.
We shared our demonstration with a notoriously inaccessible population of expert reviewers with experience as United States Department of Defense analysts.
Our findings indicate that the proposed pipeline effectively generates cards that display key events from interaction logs, facilitating the sharing of analysis progress. 
Yet, we also hear that there is a need for more prominent justifications and pattern elicitation controls to communicate analysis summaries more effectively.
The expert review highlights the potential of automated approaches in addressing the challenges of provenance information in complex domains. 
We emphasize the need for further research into provenance communication in other domains.
  %
  A free copy of this paper and all supplemental materials are available at 
  \url{https://osf.io/j4bxt}.
\end{abstract}




\maketitle

\input{sections/01-introduction.tex}

\input{sections/02-related-work.tex}
\input{sections/03-design.tex}

\input{sections/04-system.tex}
\input{sections/05-usage-scenaro.tex}
\input{sections/06-evaluation.tex}
\input{sections/07-results.tex}
\input{sections/08-discussion.tex}
\input{sections/09-conclusion.tex}

\begin{acks}
We would like to thank our expert review participants and reviewers for their feedback on this work.
We are grateful for the assistance provided by our external collaborators and their personal anecdotes regarding implementing our approach and recounting feedback from sensitive, applied environments.
\end{acks}

\bibliographystyle{ACM-Reference-Format}
\bibliography{bibliography}

\appendix

\end{document}

%% file: sections/01-introduction.tex
\section{Introduction}
Exploratory data analysis can be complex and leave room for errors.
Analytics and software support are commonly used to facilitate data exploration, but practical cases of analysis involve incomplete or messy data that still rely on human decision-making for investigation.
In real-world contexts, data analysts do not conduct work in isolation.
Understanding of data is sought for a reason, and findings need to be communicated to others---be it superiors, stakeholders, or peers.
In some cases, a new analyst might take over for another analyst, or multiple analysts might work together to share insights.
However, people think differently and adopt different strategies in open-ended analyses\cite{Alvarado2003Surviving}, and it would be unlikely for any two analysts to explore the same dataset in exactly the same way.
Consequently, understanding the data analysis work of others requires more than only the final ``findings'' for collaborators to have confidence and trust in each others' process.

Effectively capturing and sharing knowledge of the analysis process (sometimes called \textit{analytic provenance}) is a known challenge.
Self-documentation and reporting are onerous, and they distract from the primary job of analysis.
Fortunately, ample research has demonstrated the utility of automatic software logging to capture the history of the data analysis workflow.
Software logs can record the sequence of user interactions as well as the history of data accessed over time, and prior work has found that review of such provenance records substantially improves understanding of prior analyses~\cite{Ragan2015LOD}.
This approach can capture the analysis history automatically while eliminating (or greatly reducing) the need for manual human documentation, and the history can be reviewed by others to meet the needs of collaborative settings or reporting.
The downside, however, is the large scope of software logs and the large quantity of captured information.
Major challenges remain for how to transform analytic provenance information into an easily-understandable format.
While researchers have designed many visualization solutions to facilitate deep review of analytic provenance~\cite{Dunne_Henry_Riche_Lee_Metoyer_Robertson_2012, Zhao2018handoff}, our work addresses the practical need to summarize the workflow for communication.

This paper presents a unique way of processing analytic provenance  information with an emphasis on summarization for easier human interpretation. 
Interested in simplifying analysis sessions into more easily understandable chunks, our approach breaks down interaction sessions into segments, summarizes the associated data interests and interaction activity, and generates a set of cards with visually-annotated natural language summaries.

\add{While our proposed approach could be applied to many different analysis domains, we chose to focus on a single domain as a case study that demonstrates one way that our segmentation and summarization methods can be implemented.
Because} different communities and analysis domains have different priorities or preferences for summarization,
\strike{we focused our design on the intelligence analysis domain as a concrete basis for understanging user needs.}
\add{our case study focuses on a single domain.
Specifically, our design is focused on the intelligence analysis domain as, broadly speaking, intelligence analysis involves making sense of large amounts of different data types from various sources, including textual data, images, and network data, although we primarily focused on textual data for our case study.
This domain provided us with an environment that focuses on collaborative work, requiring team members to understand what their teammates have been thinking and doing.}
This paper presents our technique along with a case study in which we worked with experienced professional analysts to realistically recognize expectations and practical needs for summarizing analytic provenance.

We make the following research contributions in this work.
\begin{itemize}
    \item We propose and demonstrate a way to segment and summarize interaction histories for hand-off communication.
    \item We interview experts in the intelligence domain using our demonstration as a design artifact to prompt realistic responses.
    \item We discuss implications and suggest takeaways from the use of segmentation and summarization in real-world collaborative analysis environments.
\end{itemize}





%% file: sections/02-related-work.tex
\section{Related work} \label{sec:related}

Across many domains, exploratory data analysis does not start and end with a single person. 
Often there is a need to support collaborative work.
Communicating findings and processes can be aided by provenance, or the steps of how something develops over time.
In this section, we discuss some of the scenarios in which collaborative analysis is conducted, the tools used, and the ways others have communicated work complete.

\subsection{Collaborative Visual Analysis}
Understanding complex issues often requires deliberation and discussion with others to fully comprehend.
Heer et al.~\cite{Heer2009voyagers} were an early example to recommend a set of features that make analysis more collaborative.
Since then, there have been numerous examples that have incorporated support for team collaborations.
For example, Dunne et al.~\cite{Dunne_Henry_Riche_Lee_Metoyer_Robertson_2012} display all the events involved in analyzing archaeology data to more completely \strike{express}\add{show} investigation steps and the types of data being explored over time by multiple users.
\add{Analysts can also have a large degree of fuzzy knowledge about the datasets they work with. 
Lin et al.~\shortcite{Lin_Akbaba_Meyer_Lex_2023} conceptualize \textit{Data Hunches} as a way of recording and communicating the knowledge that the analysts bring to their data analysis process.
By allowing for scribbles directly on visualizations, they suggest guidelines for facilitating collaborative analysis.
}
\strike{Expressing}\add{Displaying} completed work has also been beneficial in the intelligence domain.
Both Goyal and Fussell~\cite{Goyal_Fussell_2016} and Zhao et al.~\cite{Zhao2018handoff} designed tools that essentially share group understanding as either bar charts or knowledge graphs to aid in communicating understanding. 
Often, in this domain, they rely on ``coverage-based'' techniques that show what kinds of data have received attention to more efficiently describe the investigation status.
For example, Chart Constellations show which data columns have received the most attention by a collection of analysts and aid in the description of what data still need exploring~\cite{Xu_etal_2018_chart}. 
Yet, understanding the order of analysis can still be helpful for communication.


\subsection{Analytic Provenance}
Collaborative visual analytics often lacks support for sequences of events because it is difficult to derive meaning from all the steps in an efficient way~\cite{Bao_2013}.
Often, in these exploratory settings, the signal representing one's final result is crowded by a cacophony of smaller events and details that would help reproduce similar results but would be impractical for others to review.
The sum of these smaller events describes analytic outcomes, behaviors, and strategies.
One solution is to analyze the event data to extract meaningful patterns and make it easier to digest; we call this analytic provenance~\cite{Xu2020Survey,ragan2015characterizing,north2011analytic}.
\add{Gadhave, Cutler, and Lex~\cite{Gadhave_Cutler_Lex_2022} proposed and reviewed a technique to capture, schematize interactions into semantic chunks from exploratory analysis.
These provenance graphs could then be reapplied to different datasets and contexts to try applying similar data filtering and preparations to different data contexts.}

Analytic provenance can be represented in multiple ways.
Often, the way provenance information is presented can dictate its use~\cite{ragan2015characterizing}.
For example, provenance data can be used for personal retrospectives of work completed in a day or more holistic representations, referring to a meta-review of perhaps an entire department.
Retrospectives are important for various contexts: including recommendations about past notes and contexts of behaviors~\cite{Shrinivasan_Gotz_Lu_2009} and even for refreshing user memories~\cite{Ragan_Goodall_Tung_2015}. 
Prior work has emphasized the use of narrative structures~\cite{Segel_Heer_2010} or designed tools that support the presentation of analysis results~\cite{Norambuena2021Narrative, Deokgun2022Storyfacets, Mathisen2019InsideInsights, Sevastjanova2021VisInReport, Gratzl2016CLUE}.
While considering the narrative structure of data presentation in a report is important, as seen in these examples, building these narratives still takes time and requires additional steps to call attention to key moments in an analysis session to prepare a presentation.
Essentially, using a narrative helps to put the information in a familiar structure. 

Essentially, Gotz and Zhou~\cite{Gotz_Zhou_2009} recognized that provenance information has a hierarchy of semantic intent. 
Provenance at the lowest level (i.e., event-based provenance) is helpful for transparency and the reproduction of tasks. Still, provenance at the highest level (i.e., task-based provenance) is helpful for communicating task completion.
Further experimental evidence was found by Block et al.~\cite{Block_Esmaeili_Ragan_Goodall_Richardson_2022}, where the higher-level provenance (coverage-based provenance) was more helpful than a list of events (event-based provenance) when handing-off a collaborative analysis task. 

Yet, questions remain regarding representing these details in different contexts and domains.
Lemieux describes the variety of purposes served by provenance in different fields and representations it can take~\cite{Lemieux_2016}.
Calling from education literature, Mayer and Pilegard~\cite{Mayer_Pilegard_2014} state that ``people learn more deeply when a multimedia message is presented in learner-paced segments rather than a continuous unit.''
This is to say that \strike{by} pacing information into chunks supports learning by reducing cognitive load~\cite{Castro2021five, Biard_Cojean_Jamet_2018_segmentation, Klingenberg2023Facilitating}.
One recommendation is to separate and share semantically meaningful pieces of the sequence and use a visual analytics tool to identify meaningful patterns~\cite{kwon2016peekquence}.
Yet these techniques are commonly focused on providing access to provenance data while still asking the user to complete the analysis.
They expect users to identify interesting features and, ultimately, complete the meta-analysis of an analysis process.
Yet, analysts only have so much time and need help parsing analysis history.

\subsection{Text Summarization} 
Textual reports commonly communicate information within multiple professions, including nurses' notes~\cite{Dracup_Morris_2008, Wayne2008Simple, Chircu2013Medication}, between rocket control shifts~\cite{Patterson_Woods_2001}, and collaborative investigations~\cite{Zhao2018handoff}. 
Reading is a standard skill set and does not require complex instructions to help users navigate information.
Written words are common across domains, with examples ranging from emails, informal text messaging, internal reporting, and writing for external publications.
Visualization systems have also begun to incorporate natural language generation to better explain data facts and charts~\cite{Srinivasan_2019_Augmenting}. 
With more advanced natural language processing (NLP) techniques (though major problems still exisit~\cite{Ji2023Hallucination}), there is the potential to not only explain a dataset but also automatically generate coherent summaries that highlight insights from an analysis~\cite{Srinivasan_Stasko_2017}.
However, in the NLP domain, the focus has been on both automated scoring functions (e.g., BLEU, ROUGE, etc.) and human evaluations to determine the agreement between a summary and the individual text document it was generated from~\cite{Fabbri2021SummEval}.


In the realm of analytic provenance, the goal of summarization is not to shorten text corpora but to communicate interaction histories.
Interaction histories are their own unique challenge because there is no ground truth ``best'' summary for an interaction history, and the interactions are often just logs of events and not text documents a sequence-to-sequence model typically expects~\cite{Sutskever_Vinyals_Le_2014}. 
Calling on narrative theories to express the sensemaking process is not new~\cite{Norambuena2021sensemaking}.
Cunningham et al.~\cite{Cunningham2019narratives} have shown how Rhetorical structure Theory can frame WC3-compliant provenance data to adjust the level of detail for different users.
Meanwhile, Chen, Xu, and Ren~\cite{Chen_Xu_Ren_2018} and Veras and Collins~\cite{Veras_Collins_2017} show how the minimum description length principle can be used to trade levels of detail for visual clutter and adjust for different kinds of users.
For this reason, our work recognizes that existing solutions can be adapted to prepare natural-sounding textual summaries tailored to the needs of any specific domain.
In the example implementation presented in this paper, we aim for simplified natural language summarization in order to focus on a small set of descriptive characteristics of analysis activity.

\add{
In the literature about collaborative sensemaking and provenance supports, it is common to incorporate aspects of narrative visualizations and presentation support features as part of visualization tools.
These techniques draw on the aspects of different narrative structures~\cite{Segel_Heer_2010} or design an environment that integrates a presentation support feature into the analysis~\cite{Deokgun2022Storyfacets, Mathisen2019InsideInsights, Gratzl2016CLUE}.
While considering the narrative structure of data presentation in a report is an important consideration, as seen in these examples, building these narratives still takes time and requires additional steps to call attention to key moments in an analysis session to prepare a presentation~\cite{Norambuena2021Narrative, Deokgun2022Storyfacets, Mathisen2019InsideInsights, Sevastjanova2021VisInReport, Gratzl2016CLUE}.
To help address these time concerns, there is value in exploring automated storytelling techniques to generate narratives for analytic provenance summarizations.}


%% file: sections/03-design.tex

\section{Design Overview}
In this work, we offer a technique to segment and summarize interaction histories to support information hand-offs, such as for analyst-to-analyst and analyst-to-manager communications.
In prior work, we see techniques focusing on types of data interacted with~\cite{Sarvghad2017Coverage}, but there are benefits to transparency, accountability, and trust-building when the process is also emphasized~\cite{Block_Esmaeili_Ragan_Goodall_Richardson_2022, ragan2015characterizing, Groth_Moreau_2013PROV}.
\strike{These considerations motivate the methodology described in this paper.}
\add{In this work, we want to shed light on what features are helpful to describe a process without relying on glyph-based, high-level coverage or a comprehensive, detailed list of all the steps taken to solve a problem.}


\subsection{Design Rationale and Goals}\label{sec:goals}


\strike{It is natural human behavior to break large, complex, overwhelming problems down into manageable chunks to apply our mind's often serial processes.} 
\add{In the handoff of work between team members, it's important to consider cognitive load theory~\cite{Young_tenCate_OSullivan_Irby_2016}.}
\add{Cognitive Load proposes that there is a limit to the available working memory of an individual~\cite{vanBruggen_Kirschner_Jochems_2002}, and the ability to comprehend is related to how information is grouped together as smaller, more familiar units (i.e., schematized)~\cite{Sweller_Ayres_Kalyuga_2011}.}
\add{But often there are hundreds of events to communicate, and doing so without overextending the receiver's cognitive load is not easy~\cite{Young_tenCate_OSullivan_Irby_2016}.}
People can perceive more, have better memory, and communicate events better when that information is segmented~\cite{Zacks_Tversky_Iyer_2001}.
Psychological research has shown how \strike{the} concretizing \strike{of} events into smaller segments helps in understanding those events too~\cite{Zacks_Tversky_2001, Zacks_Swallow_2007}.
Evidence of the benefits of segmentation can also be found in human-computer interaction research.
They show how providing aggregations of interaction events and dynamic binning of information can be beneficial when reviewing temporally relevant event sequences~\cite{Wang2009Temporal}.
We also see design recommendations for explaining data using story backbones and layouts found in comic books to aid in communication~\cite{Bach2018datacomics}.
Segmentation is a natural way that humans perceive episodic behaviors, so \add{we theorize that} having a technique that can separate event sequences into appropriate chunks would be useful.

\strike{At the same time, there are also calls for \strike{meaningful} summarization of interaction information.
On the one hand, detailed representations of provenance information can aid in visual analytics when the goal is to describe steps transparently.}
\add{Interaction information is typically captured in full detail, but when everything is captured, it takes people time to parse and identify what they are looking for.
The default, comprehensive capture of provenance information is helpful when the goal is to describe steps transparently.}
To illustrate this point, consider SenseMap~\cite{Nguyen2016SenseMap}, a browser extension that captures and displays \strike{pages} visited \add{web pages} as an expanding tree.
This tool can capture an exhaustive list of page visits, but as more pages are visited, the tree continues to expand and can become overwhelming to review.
These types of tools---which aim to capture and facilitate a review of all the behaviors in a session---are especially helpful for transparency purposes or personal reflection; however, to be useful for other users who are trying to understand the gist or overall flow of the analysis, the complete history is often overwhelming. 
Simplification\add{ or summarization should lessen the required cognitive load and} is a priority for effective communication and integration into practical work styles. 

Other methods propose a succinct overview of provenance information by showing what information was interacted with over time.
This focus on what was covered can \strike{be helpful for explaining}\add{help explain} work done by researchers~\cite{Liu2020Paths}, showing what information has already been analysed~\cite{Sarvghad2017Coverage}, or encoding visit history in the visualization's default view~\cite{Feng_Deng_Peck_Harrison_2017}.
Some techniques, like Zhao et al.~\cite{Zhao2018handoff}, include an interactive timeline with a knowledge graph that shows what a team knows over time.
\strike{Yet, }When comparing provenance visualization techniques for hand-off communication, Block et al.~\cite{Block_Esmaeili_Ragan_Goodall_Richardson_2022} found that providing summaries can be more efficient because providing too much detail creates a new, secondary sensemaking task for the user to try to analyze the tasks of a prior user.
\add{
Managers are not expected to complete the same work as their team and are more concerned with work coordination and worker performance. 
One of the goals of analytic provenance is the ability to identify and extract meaningful insights from interactions over time.
But it is important to consider that different users will have different goals for provenance information~\cite{ragan2015characterizing,Fisher_Counts_Kittur_2012} and will likely need the data summarized at different levels of detail.
As described by Brehmer and Kosara~\cite{Brehmer_Kosara_2021} the level of collaboration and audience types should influence what controls are available to an audience and the expectations for synchronous and asynchronous collaboration.
Li, Luther, and North~\cite{Li_Luther_North_2018} distributed the analytic process among crowdworkers, defined five key sensemaking steps, and a custom interface for each stage of the analysis process to provide enough ``context slices'' to allow crowds to solve mysteries.
In the analysis process, it is important to have some differences in how information is presented to make it easier to comprehend.
}
\strike{While there are places for deeper, more holistic representations of the progression of analysis, we focus our method on representations that inherently simplify analysis workflows.}

\strike{Summarization is not only helpful for collaborators (i.e., people working toward the same goals), but also for managers who supervise multiple individuals.
Managers are not expected to complete the same work as their team and are more concerned with work coordination and worker performance. 
Summarization helps provide information at a high level, thus making it human-understandable.
Adjusting what is summarized at different levels of detail can be especially helpful for both managers and collaborators.}
\strike{Our goal, as we describe next, is to communicate an overview of the flow of analysis and progression of topics/items of interest in summarized chunks at different levels of detail to help reduce overwhelm and cognitive demands. }
\add{Based on the literature, we believe interaction information should balance the following goals: minimize the cognitive load, compartmentalize chunks of information, and provide sufficient summaries of completed work.}
\add{While the best representation of provenance information will likely be application-dependent, we examine a specific technique from the lenses of intelligence analysis working with textual data while showing some history of the analytic process to aid in work handoff.}

The \strike{three}\add{desired} goals for our overview technique are to:
\begin{enumerate}[label=\textbf{G\arabic*.}]
    \item Temporally segment analysis sessions into digestible chunks by using familiar reporting schemes (i.e., use natural language, lists, and simple data visualizations). 
    \item Summarize  interaction logs to communicate processes efficiently with an appropriate amount of meaningful semantic information.
    \item Enable identification of different analyst behaviors and patterns, as would be desirable for collaboration, managerial review, and mentoring purposes.
\end{enumerate}
\add{
Based on the literature, we believe that designing a technique that segments interactions, summarizes analytic behaviors, and elicits patterns automatically would help reduce cognitive load and make information handoff more efficient and less error-prone.
Integrating common web technologies, we created a design artifact, shared the technique with intelligence experts, and used a task-based interview to understand better what features may be helpful to users with frequent information hand-offs and lots of information to manage.
}

%% file: sections/04-system.tex
\begin{figure*}
      \centering
  \includegraphics[width=\linewidth]{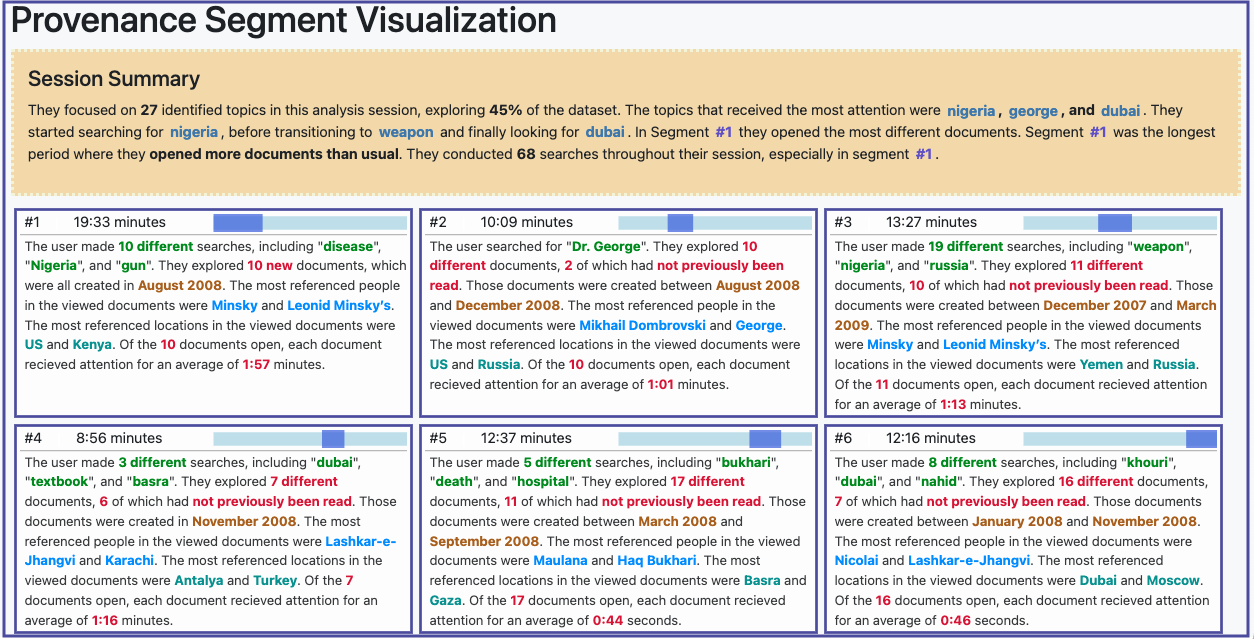}
  \caption{The interface we designed to help communicate analytic provenance information to new users.
   Notice that it contains an overarching summary (in yellow) and individual summaries (the set of cards) for segments of time to help explain the analysis process.
   In this paper, we present the visualization approach and demonstrate the automated technique on five different datasets. 
   }
   \label{fig:interface}
\end{figure*}

\section{System Design}\label{sec:system}
The contributions of this paper are essentially twofold. 
We recommend a methodology for segmenting and summarizing analysis sessions and demonstrate the applicability of our technique by collecting feedback from experts from the intelligence domain.
In this section, we describe the specific components required for our demonstration and the visual interface.
We re-iterate that the specifics of our approach are not strictly applicable to the intelligence domain but rather serve as a laboratory for understanding the impacts of segmentation and summarization for communication.
\add{This design simply serves as a case study to provide one example of how our proposed approach could look when implemented.}
We recommend further exploration into \add{other} domain-specific designs.

\add{Due to very restricted availability with domain experts in the intelligence analysis field, we could not perform rigorous evaluation to establish formal user requirements before beginning system development.
We were limited to a few informal conversations to understand the context of their working environment at first, and were later able to meet with them in more formal, long-form sessions for evaluation of our designs.}

\subsection{Visual Interface}\label{sec:interface}
To support \strike{the} \add{our recommended} design goals for workflow summarization, our approach generates a visual interface of one analyst's analysis history in a web application.
The interface essentially consists of two parts.
Aside from a basic control panel for selecting parameters to display, there is an overview brief at the top and a series of cards showing interaction information at different points in time.
We describe each component in the following subsections.


\subsubsection{Session Summary Component}

In the yellow box at the top are a series of sentences describing stand-out periods or key features of an analysis history that serves design \textbf{G2} by providing a broad overview of session highlights and calling attention to segments of interest.
\add{As described by Sarvghad et al.~\cite{Sarvghad2017Coverage}, providing an overview is helpful for establishing context and common ground~\cite{Robinson_2008}.}
In the visualization we present, this overview is calculated outside the segmentation algorithm and dynamically adjusts the content based on the active session.
While it would be possible to call on large language models like ChatGPT~\cite{gpt} to generate this overview, their output can be hard to control, inaccurate, and does not provide obvious hooks for applying interaction supports\add{\cite{Ji_Lee_Frieske_Yu_Su_Xu_Ishii_Bang_Madotto_Fung_2023}}. 
For this reason, we rely on template sentences for more control and to allow some terms to respond to hover events in the interface.
The use of template sentences is partially inspired by Packer and Moreau~\cite{Packer2015templating}. 
They suggested using sentence stubs as an explanation interface for W3C provenance data.
Ultimately, their approach was a simple translation of relationships in a provenance graph to a set of sentences with little support for tying different sentences together.
To fill in the gaps in the sentences, we calculate a series of statistics, including the most frequent search terms, the number of keywords, the percent of the dataset reviewed, and the average rate of interaction across the entire session.
Since these statistics are calculated after the segmentation is complete, we can also calculate superlatives for each segment and call attention to them to better guide the user's exploration (e.g., ``Segment \#1 was the longest period'').
\begin{figure}
    \centering
    \includegraphics[width=0.65\linewidth]{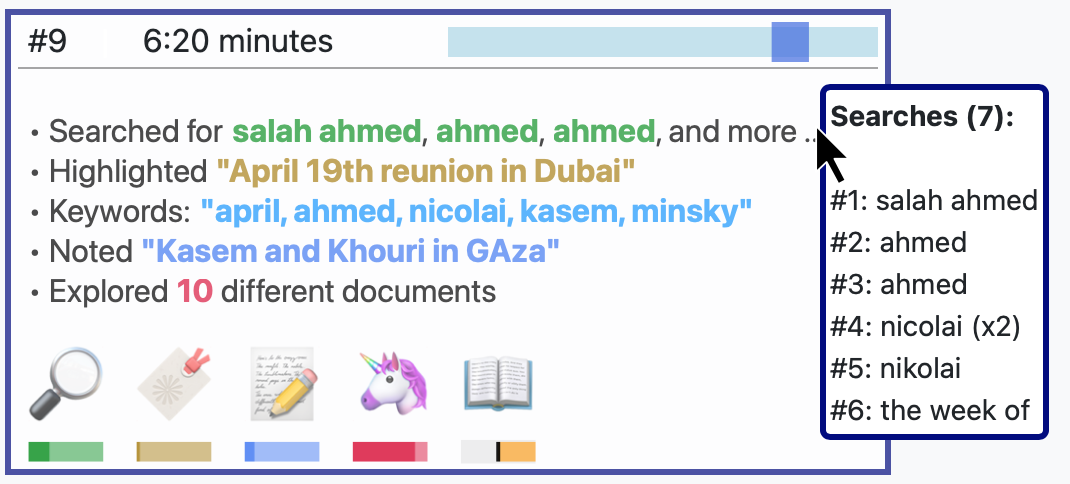}
    \caption{A single card in list mode shows the information from a segment more completely, especially to aid debugging or for more detailed review or interaction provenance.
    The tooltips help show the various data more thoroughly.
    For example, we can see the list of searches completed in this segment when hovering over the list.
    }
    \label{fig:listmode}
\end{figure}
\subsubsection{Card Summary View}
Below the overview is a set of cards summarizing segments of a session. 
\add{Although, Goyal and Fussel~\cite{Goyal_Fussell_2016}, found that cards were less helpful than visualizations for communicating work completed in the handoff of information, their approach was focused on analysts with shared work.
Our focus in this work is more related to reporting provenance information to audiences with different levels of expertise and adjusting the display of provenance information. We choose to use textual cards as textual intelligence briefs are commonly used in the intelligence community~\cite{block2023Preliminary}. }
The cards go in order of time, starting at the top left, and each card has a blue time bar at the top representing what period of the session is summarized.
The goal of these cards is to summarize stages of analysis.
Our visual design essentially makes the entire sequence of analysis visible at once.
We encode different types of information using different colors for scanability (e.g., green for search information, red for document open events).
This design makes it possible to see key items of interest over the progression of the data exploration (\textbf{G3}). 
To help users parse an analysis session's content, we separated an interaction history into smaller segments and summarized them.
These 11 segments were pre-processed (as described in Section~\ref{sec:algorithm}) and rendered to the screen to show various information like the searches complete, documents open, and entities related to documents open.
Users can look for similarities among cards to get a feel for the most salient points to focused on, and looking at the differences gives a sense of the evolution of thought.

There are two modes to the interface, and we have a toggle to adjust the view of the data. 
In \textit{prose mode} (see Figure~\ref{fig:interface}), users can read a natural language summary of the events in a segment, whereas, in \textit{list mode} (see Figure~\ref{fig:listmode}), the data is provided as a list with tooltips to show more details as you mouse over items.
For example, the keywords are extracted from which documents were opened during a particular time segment. 
In the list mode, we also provide small word-size graphics, with the aim to support user review of the interactions and minimizing cognitive load~\cite{Beck_Weiskopf_2017,Sweller_Ayres_Kalyuga_2011}. 
The paragraphs tell the story, and the list view shows all the data and requires mouse-over interactions to reveal details.

\subsection{Data Processing}\label{sec:algorithm} 

At a high level, our technique uses documents and interactions on those documents to generate a series of segments and summarize an investigation.
We want to segment analysis sessions into chunks automatically to make them more digestible (\textbf{G1}) and then summarize those segments to explain the story of someone's activity briefly and quickly (\textbf{G2}).
To support information hand-off and the identification of different analysis styles (\textbf{G3)}, we chose a methodology that would convert semantically poor interaction logs to something more familiar and digestible (\textbf{G1}): Textual Reports. 
\begin{figure}
 \centering
  \includegraphics[width=\linewidth]{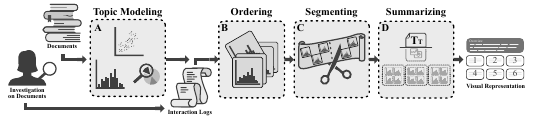}
  \caption{An abstraction of our segmentation and summarization methodology for interaction histories in analysis scenarios.
   Essentially, our technique uses A) topic models of the underlying dataset and B) the order of interactions with these documents to identify C) key breakpoints to segment an investigation along.
   D) We summarize each segment as textual cards to tell the story efficiently}.
  \label{fig:teaser}
\end{figure}
At its heart, our method (As seen in Fig.~\ref{fig:teaser}) is essentially four phases.
A) vectorization of document content, B) synchronization of document visits with the order from interaction histories, C) segmenting the history based on document content interacted with over time, and finally, D) summarization for each segment of identified time.
At a high level, we describe this technique and its potential benefits for one community.
Yet, every domain interested in this approach will have unique data challenges.
Therefore, we demonstrate the fundamentals of the technique by implementing a basic rendition of our pipeline without fine-tuning or selecting the best components, as these will adapt to every scenario.
Below we describe the steps we took to prepare a non-specific version of this conceptual technique.

\subsubsection{Preparing Session Overview}
To begin, we define a set of metrics to describe the entire interaction history and support the overview interface component.
As part of the pre-processing phases, an object is defined that describe some superlatives for a user's interaction session, like the percent of the dataset explored, or the total number of searches complete.
This set of superlatives can then be updated as segmentation occurs (see Section~\ref{sec:segmenting}) to describe key behaviors, identify anomalous segments or call attention to topic changes over time.

\subsubsection{Preprocessing for Segmentation}\label{sec:segmenting}
There are many possible ways of preparing the data for segmentation and summarization.
In our specific implementation, we calculate a Term Frequency-Inverse Document Frequency (TF-IDF) vector for each document.
We then loop over the events in an interaction log and replace any document visit with its TF-IDF vector.
For any semantically meaningful events (e.g., query text, highlighted information, notes taken, etc.), we consider the textual content of these interactions as their own documents.
Once the pre-processing has been completed, we have a vectorized form of the document and interaction content over time.
To divide these interaction histories into segments, we use iterative binary segmentation, an offline-time change point detection algorithm~\cite{Truong2020change}.
At each iteration, the algorithm selects a point in time that maximizes the difference between the aggregated vectors of the two new segments that would be created by slicing the existing segment at that point.
This is repeated iteratively until either a set number of segments are created or until there is no new segmentation point that would result in sufficiently different segments.
Based on internal testing, we chose 10 slices resulting in 11 segments for this basic implementation. 
The choice of segmentation algorithm and its parameters may vary greatly between applications.
Our specific selections merely serve as an example.

\subsubsection{Preparing Summarization Cards}
Once the interaction histories are segmented, we can generate summarizations.
Multiple ways exsist for summarizing content including the use of large language models or a human in the loop approach.
In our specific implementation, to summarize the content of the documents and textual user interactions within each segment, we identify the top words from the aggregated vectors to display as keywords.
Additionally, we use named entity recognition from a spaCy English trained pipeline\footnote{spaCy  available at \url{https://spacy.io/models/en\#en_core_web_md}} to extract the most referenced people and locations from the documents interacted with in each segment.
These are then prepared for the visualization component (described in Section~\ref{sec:interface}).
In our case, we choose to represent these patterns as text because it does not rely on visualization literacy and can explain potential patterns more explicitly without relying on taxing user exploration and analysis.
Ultimately, we demonstrate this algorithm with a basic set of components to better understand what recommendations would benefit a tool used to summarize work in the wild.
The code used to process and display our technique can be found at 
\url{https://github.com/brettbenda/ProvenanceSummaries}

%% file: sections/05-usage-scenaro.tex
\section{Usage Scenario}
Our tool allows a user to quickly grasp the content and focus of a previous work session. 
This can be used to refresh oneself on prior work, ease the hand-off of projects, or review the work of a peer or employee. 
Here, we present usage scenarios for each of these cases to contextualize the features of the tool.

Consider Jack, a researcher who is conducting a literature review along with his colleague Jill, under the supervision of their manager, Madison. 
Jack spends one work session accessing dozens of documents that contain common threads, such as discussion topics or references to items or figures.
During the morning work session, Jack searches for key terms, highlights items of interest, and takes notes on his conclusions and later questions before taking a break for lunch.

When Jack returns, he is able to review his research session by exploring the list view.
Here, his sequence of actions is more transparent.
The interface adapts to the number of interactions from the morning, and without having to scroll the page, he can recall the list of searches conducted as well as where he wrote more notes.
This summary of his session serves to quickly remind him of the work he has already done and reflect on other directions he can research in the afternoon.  
This demonstrates a critical value of our recommended pipeline: by showing analysis processes as segmented cards, familiarization with past work is easier to do, because the analysis is broken into stages and summarized.

The next day, Jack is unavailable, and Jill is tasked with continuing the literature review. 
To pick up where Jack left off, Jill would need to understand, at a high-level, the status of their shared project and any potential areas where she could contribute.
Without a tool like the one we describe in this work, Jack would have had to keep notes about his investigation status another way (likely with manually written notes).
Seeing Jack's investigation session segmented and summarized in our tool is helpful for inferring the main topics already explored and the general flow of logic.
By reading the summary at the top, Jill can see what topics received the most attention from her colleague and the segments with significant activities.
Looking at the summary cards, she can view the various segments that made up Jack's work sessions, referring to the keywords and entities involved with the documents accessed in each of the segments.
Even without referring to the more detailed view, she is able to familiarize herself with Jack's work in a few minutes and set a trajectory for herself to continue the literature review.
This shows another important aspect of our technique: by having different levels of summarization, individuals interested in understanding the process more holistically can do so without zooming into all the specific events.

When the literature review is complete, Madison is tasked with reviewing the work and reporting on the performance of her team. 
Using the segment visualization tool, she is able to see the processes used to arrive at the final result. 
This can inform the edits she makes. 
For example, she can tell what terms were commonly searched for by comparing the two session reports and more confidently report on how thorough she feels her team has been in their review, all without having to ask Jack or Jill to prepare a report or arrange a one-on-one meeting.
Perhaps more crucially, this tool can also inform future training.
Looking at the steps taken by her two workers gives her insight into the work process of Jack and Jill, and she could use this information to offer constructive feedback not just on their results, but on their process and techniques.
This scene represents how the segmentation and summarization of past work can help automate reporting, especially reports describing business processes.








%% file: sections/06-evaluation.tex
\section{Case Studies with Intelligence Analysis Datasets}

The inherent nature of provenance communication is that it must match the needs of the specific audience or users~\cite{madanagopal2019analytic}, so the specifics for the implementation of our proposed segmentation method must be tailored to a particular purpose.
As a demonstration of our technique, we conducted a case study in the domain of intelligence analysis.
Of the large number of domains interested in provenance support~\cite{Xu2020Survey}, the intelligence community relies heavily on reporting to share information with decision and policy makers~\cite{SigintAnnex2021}.
We look at a scenario of intelligence work using provenance summaries of analysis for \textit{pass-downs}, which refers to one analyst communicating complete and outstanding work to a new analyst.
In practice, pass-downs are often static, manually generated text documents, and interaction and automation could save analysts time and help avoid mistakes.
There are even recent efforts to generate tailored information reports for analysts of many kinds~\cite{Kershaw_2022}.
This fertile context of multiple analysts collaborating or managers reviewing analyst teams makes an ideal focus for reviewing our tool.

While our external collaborators in the intelligence domain could not share the specifics of the datasets and tools they currently use, we could identify proxy analysis interaction log datasets completed on both fictional and real-world data.
These logs shared a similar structure to the interaction information captured by intelligence tradecraft and tools.
We discuss the five test datasets and the process of adapting our technique accordingly.

\subsubsection{Interaction Logs with \add{three} Synthetic Text Datasets}\label{sec:dataset_descriptions}

Three of the provenance datasets were based on interaction logs from sample analysis sessions conducted with fictitious text document data~\cite{Mohseni2018datasets}.
The VAST Challenge~\footnote{The most recent IEEE VAST Symposium was in 2023: \url{https://vast-challenge.github.io/2023/about.html}} is an annual competition within the visualization community designed to provide participants with an opportunity to work on real-world data sets and tackle challenging problems in visual analytics.
It is important to note that all of these documents are fabricated and designed to mimic the kinds of textual reports seen by intelligence operators.
The documents encountered often include newspaper clips, email correspondence, bank transactions, and more, including the occasional misspelling of names and incompleteness of data experienced by intelligence operators.
Based on previous VAST Challenge datasets, prior records of provenance log datasets from Mohseni et al.~\cite{Mohseni2018datasets} recorded the interaction history from students using an analysis tool to review three separate textual data associated with the VAST mini-challenge.
Each of the three sample datasets considers the interactions complete on 102--158 text documents.
The interaction logs from the datasets comprised interactions completed within a visual analysis tool for searching and viewing text documents.
This tool consisted of an explorer with moveable windowed documents that allowed users to read, search for exact terms, highlight text, and write their own notes.
In total, logs are provided from twenty-four participant sessions using this tool for about 150 minutes each to solve one of the three mysteries (i.e., eight participants per mystery).
\add{A copy of these datasets and the code needed to generate our interface are available at the OSF link here} 
 \url{https://osf.io/j4bxt}

\subsubsection{Interaction Logs with Language Analysis of Conversation Records} \label{sec:nixon_tapes_interactions_dataset}
To further demonstrate our technique's applicability, we also applied our approach to the interactions completed on a dataset collected independently by external collaborators.
In this case, they had interactions from participants mimicking the work completed by language analysts in an unclassified setting.
They developed an independent analysis tool that helped find audio recordings and imperfect transcripts from a publicly available dataset of former United States President Richard Nixon's phone recordings from his time in office\footnote{These audio samples are cataloged and available to the public at \url{http://nixontapes.org/transcripts.html} though the test dataset was processed and generated by the external collaborator's systems.}.
To better understand how individuals work with large corpora of audio recordings, they asked participants to identify when events took place, decode code words and resolve different data discrepancies while tracking their interactions.
The tool had an advanced search function allowing for not only keyword identification, but also could be refined by certain speakers, tape recording locations, and ranges of dates.

For this dataset, eight participants conducted the sample analysis.
Many participants had training as professional analysts, but some had little to no analysis experience.
Their analysis tool logged interactions for any searches completed, any transcripts opened, and when an audio file was played.
The transcripts were exceptionally long since many tapes contained the beginnings and ends of multiple conversations upwards of two hours in length.
For context, participants commonly performed a search, reviewed the list of results, opened a transcript, and spent a long time playing back the associated audio to find the information of interest.
Therefore, the interaction logs included far fewer and less frequent interactions than the fabricated datasets described prior.
This specific context contrasts with the earlier three datasets because it referred to long audio files with often incorrect transcription instead of only relying on text and misspellings.
We showed that with a custom pre-processing step, we could translate the interactions into a format expected by our tool for summarization. 
This implementation helps show that our pipeline can be adapted and fine-tuned to accept interaction logs from existing tools.

\subsubsection{Operational Intelligence Dataset}\label{sec:classifiedDescription}

A collaborator in the specialized domain adapted our software for the summarization approach to their own dataset based on a real-world intelligence analysis setting. 
While the level of detail we can describe the dataset is limited due to its sensitive nature, we describe the main points important for understanding the similarity to and differences from the previously described data sets.
The implementation used an existing dataset designed for evaluating the search performance of intelligence tools and compliance checking among intelligence community offices to improve internal tool development and inform analysis strategies.
The interaction dataset was previously logged with an application deployed on a system to look at query metadata for query compliance, a formally defined requirement of the intelligence community~\cite{SigintAnnex2021}.
The external collaborator selected a single office's 24-hour search activity to generate a static demonstration for their discussion based on actions involving queries for information.
This consisted of a small collection of people making queries with rich metadata information about what was being searched for and the associated justification. 
The query logs also included various types of metadata recorded for compliance purposes. 
For example, they included timestamps, document sources, and information relevant to access restrictions.
In practice, such metadata is often captured but not directly visible for review.
For the case study, they included the query string, the number of results (yet, not the resulting documents associated with a query), the date and time of the query, and the designator requesting the information.

The approach was adapted with minimal issues for creating brief collaborative summaries. 
Due to the sensitive nature of the data context, the data did not include links to the underlying corpus of documents being searched in the interaction logs.
This made automatic segmentation difficult since the binary segmentation (described in Section~\ref{sec:algorithm}) relies on the computed difference in document topics over time.
Instead, our external collaborator used only the query metadata for segmentation and could account for data similarity or differences across segments.
This demonstrates a practical challenge in summarization for analysis cases where data privacy or access restrictions might be in place.
However, the approach can be adapted to operate on a subset of available data features to circumvent such limitations.

\section{Evaluation}

 We evaluate our segmentation and summarization of individuals' behaviors  with experts in the intelligence domain.
 We discuss the user experiences and qualitative feedback we received from providing our technique and its demonstration as a design artifact.
 Through this evaluation, we aim to gain insights into the effectiveness and utility of our approach in supporting investigators' decision-making processes.

\subsection{Expert Review}

To evaluate the technique and implementation, we conducted an expert review with five domain experts in intelligence analysis.
Historically, these expert users are challenging to recruit due to the restrictions on their available time given the size of their workload\cite{block2023Preliminary, Johnston_2005}.
The experts had several years of experience within the intelligence community and dealt with the passing of intelligence information throughout their careers. 
We identified this group because of their experience with passing information and the similarity to the kind of interactions available in the test datasets (see Section~\ref{sec:dataset_descriptions}).
Three of the five experts had experience managing other analysts, and they all served in many capacities, including a watch-floor officer, mission manager, and discovery analyst. 
The goal was to understand if and how domain experts could use provenance summarizations to understand the flow of an analysis completed by another person.

\begin{figure}
    \centering
    \includegraphics[width=\linewidth]{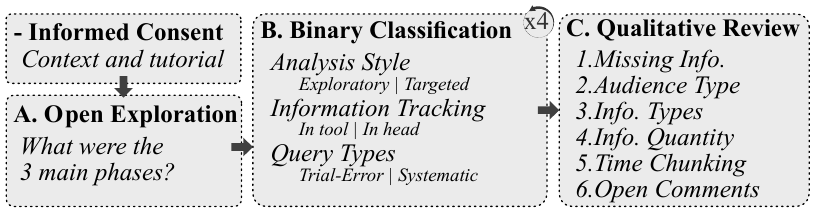}
    \caption{A visual representation of the phases of the expert review used to evaluate the visualization and automatic summarization technique.
    A) Participants openly explored the interface with think-aloud. They then
    B) classified different analysis sessions before finally being
    C) asked questions to better understand their preferences. 
    }
    \label{fig:procedure}
\end{figure}

\subsubsection{Method}
This study was approved by our organization's Institutional Review Board (IRB).
This study consisted of a semi-structured interview with domain experts in intelligence analysis.
The study had three phases (see Figure~\ref{fig:procedure}).
After receiving informed consent and describing the scenario of use, we asked participants to inspect the summarization for one person's interactions to get more comfortable with the interface and ask questions about the available features.
Participants reviewed and explored the sample analyses for approximately 10 minutes.

Next, participants conducted an analysis review task that required interpreting other analysts' strategies.
For this task, participants were asked to classify each analysis record according to three analysis styles based on previous discussions with domain experts in intelligence.
Participants were provided a set of definitions for different analysis behaviors.
They then reviewed and classified four sample analysis records in the tool (they reviewed one at a time with order randomized for each participant).
First, participants had to determine whether the analysis was either \textit{exploratory}, where it involves a more broad review of many different topics, or more \textit{targeted}, where it focuses on a set of specific kinds of information or a specific target.
Next, participants determined how information was tracked during an analysis session. 
Was it in the analysis \textit{tool}, where they used the document viewer tool to leave records of their interests and findings, or in their \textit{head}, where they left little record of their analysis in the tool until they were more confident about documenting findings?
Finally, participants decided on query behaviors, determining if the searches were more \textit{trial-and-error}, with bursts of activity and ad-hoc search frequency, or \textit{systematic} and with more focus on what they searched for and the frequency how they used searches.
Once they had classified the styles, we asked participants to rank the challenge of each classification task.

Finally, we asked participants to comment on the features we wanted to better understand.
We asked about the information they would expect to see to better understand user behaviors.
We also asked them to consider the intended audience of the tool and the modifications they would make for two likely use cases for provenance information (i.e., collaborating peers and team managers).
We then asked about the quantities of information on a card and the chunking among the cards before opening up for any final generic comments as well.
All of these later questions were considered through the lens of different audience types to better understand how peers' and managers' needs for provenance may differ.
We discuss the details of user responses in Section~\ref{sec:quantitativeResults} and Section~\ref{sec:qualitativeResults}.

\subsubsection{Analysis Review Results} \label{sec:quantitativeResults}

With the intention of exposing our experts to ways they could use our tool's features in a meta-review, we asked analysts to classify a session's analysis styles.
We asked for binary classifications for three analyst styles: analysis type, information tracking, and query type.
Experts used our interface to determine if a session was more exploratory or targeted, tracked information in the tool or in one's head, and if the searches were more ad-hoc or systematic.
We also asked participants to rank the difficulty of each classification.
Ground truth was classified by one author, who spent the most time reviewing the test datasets. 
The author selected four sessions from one dataset with an equal mix of each binary quality (i.e., 2 sessions were exploratory, 2 were targeted, 2 tracked information in the tool, etc.)
We calculate a performance score by comparing the classifications made by participants in their short 10-minute introduction to our tool compared to the ground truth determined by the author.

Overall performance was high at 76.6\%, signaling that our interface could help identify these qualities.
Performance was higher than random guessing (50\%) but clearly was still challenging.
Looking across the style categories, we see that the highest score relates to \textit{information tracking} classification (90\% correct).
This high score was expected because the interface directly lists highlights and notes made during a session.
It is likely that participants could use this feature to easily determine a classification.
The lack of a perfect score can be attributed to one session that only used highlighting and did not leave notes. 
There was ambiguity in our definition as to whether this counted as using the tool to track information.
Participants also scored highly (80\%) on the \textit{analysis type} classification, though participants consistently considered this the most difficult to determine.
Our participants said that classifying the entire session into a single category was challenging because, often, a single session can have periods of both types of activity.
Finally, analysis of \textit{query behavior} was also considered hard to determine by participants and also had the most errors with a score of 60\% correct compared to the ground truth.
While better than guessing, this does suggest that our definition for query types may have been too loose, or the representation of analysis with summary segments might make it difficult to process details of specific action behaviors.
We expected analysts to look at the frequency of search events across the session to determine if the queries were systematically dissecting a topic over time or occurring in bursts.
While this information was available to the participant, it required them to inspect the number of searches conducted in each segment of the session and determine if any segments had bursts of unusual search activity.
Ultimately, these performance scores suggest that the segmentation of interaction behaviors and their associated summaries are helpful for recognizing analysis behaviors and helped inspire further conversation about what would be beneficial from a methodology that segmented and summarized interaction history.

\subsubsection{Qualitative Feedback}\label{sec:qualitativeResults}
From our expert reviews, one of the first things our interface was recognized for was its ability to inspire questions about meta-analysis in general.
All of the experts were \strike{visually} excited by the tool's potential and emphasized the benefits provenance visualization could have across their organization.
Experts mentioned a more diverse set of audience members that could benefit from the tool than we considered initially, including higher-level management, analysis trainers, trade-craft designers, team-level management, collaboration, and even self reminders too.
This shows one anecdote into how our example helped to elicit more creative thoughts and directions for provenance information.

Looking more closely at the information on the interface, we see universal agreement that interactions with semantic information (e.g., search terms and note content) were considered especially insightful for understanding what was happening (\textbf{G2}). 
We also heard that the tool overemphasized the \textit{steps} of the analysis process and wanted to focus more on the \textit{why} for the progression of the analysis.
That is, some analysts were less interested in the details of how much time and how many documents, and they were more interested in following the flow of data topics\add{ and the rationale of each step}.
Specifically, two experts stated disinterest in the number of documents opened and the average length of time spent on each document in the textual prose. 
For further clarifying rationale, though it is possible to match behavior patterns to user intentions~\cite{Yan2021Tessera}, automatically abstracting user intention without a sense of ground truth (e.g., think-alouds) will always be imperfect.
In line with the common mantra of \textit{overview first, zoom and filter, details on demand}~\cite{shneiderman1996eyes}, one expert stated, ``I want a higher level of summarization and the ability to drill down [into more detail].''
We note that our demonstration and review focused on collecting feedback on the summarization view, though enabling link-following to access associated source information is an obvious, required feature.

Additionally, three experts looked at the text in the prose mode and mentioned that they assumed more variability for the types of information communicated in the text.
While many appreciated that the interface explained steps in prose, they noted interest in seeing more variety, unique patterns in a period of time, or details not seen elsewhere.
We set out to use text for our technique because it mimics a familiar reporting structure (\textbf{G1}) in the intelligence community. 
Still, expert feedback helps show that there is nuance to how text should be used, and the deployment for any real use case would require a more tailored adaptation of summarizations.

Related to the textual component, all experts found the full session summary to be useful for getting a feel for what aspects to pay attention to and the key features of the session (\textbf{G2}).
Yet, while some of our questions could be answered with the information in the summary section, we saw that most experts relied almost exclusively on the list mode to answer our questions.
The list mode provided details on demand with tooltips and had information spaced consistently, making scanning easier.
We believe this is because the experts were all trained analysts and wanted the most amount of detail to answer our questions.
Because the experts spent the majority of their time reviewing the cards, eventually, they all asked about how the segmentation was being computed.
As one expert put it, the ``cards emphasize the idea that each segment is distinct and different from the others.''
Echoed by others, essentially, our experts recommended explicit representation of the reason for segmenting one time segment from the others, and this reasoning must be explained more explicitly.
Segmentation algorithms with simple rules are easy to explain and would likely be better for episodic summaries (\textbf{G1}, \textbf{G2}). 
Thus, this helps shed insights into the kinds of segmentation techniques that should be considered when presenting interactions as a series of cards.

Following the expert review, we note two major changes we made to the tool.
First, we implemented a slider to adjust the number of cards to segment the analysis session into, thus allowing the reviewer the choice of the granularity for the amount of temporal segmentation and summarization.
Second, to help emphasize the flow and connection of key data topics throughout the analysis, we added a brush-and-link feature that helps to reveal other segments that contain the same keyword when mousing over the cards.

\subsection{Review in Operational Intelligence Context}

In addition to the semi-structured expert reviews  described above, we also include discussions about the feedback from demonstration of the prototype adapted to specific intelligence case.
As expressed by our external intelligence expert collaborators, many tools developed through academic research are limited to proxy data without demonstration on the actual real data of interest to specialized groups.
This collaboration intentionally confronts this problem by bridging it to a similar problem with sensitive domain data.
For this component of the case study, someone external to the core research team used our technique and the software to process their own sensitive dataset from an applied analysis setting (described in Section~\ref{sec:classifiedDescription}).
After applying the technique and generating sample analysis summaries, two associates (also external to the authors and core research team) demonstrated the output to additional experts to gather additional feedback.

The external collaborator held two informal occasions to collect feedback on the approach.
First, there was a demo session where an external collaborator provided a live demonstration and answered more in-depth questions with four people. 
Second, there was a 1 hour and 45-minute group meeting with 10 individuals.
Both occasions invited lots of interest from individuals focused on operations in general, including general technologists, technical analysts, and language analysts. 
Generally, they did not have specific opinions about the features, but the idea of the meta-analysis was seen as helpful as it does not have much support currently. 
In a later meeting, we met with this external collaborator and asked them to recall what was the experience of adapting our technique and also describe reactions to the tool.


 The applied version of the tool reviewed in the sensitive intelligence context was well received when presented to domain analysts, team managers, and upper-level management.
 The external collaborator said, ``The basic elements happily surprised the people on the high side.'' 
 The interface inspired more questions about what was possible to learn from the provenance records.
 For example, attendees considered how access to visualized workflows might influence analyst training.
 The ability to see workflows invited suggestions for detecting different analysis styles.
 As one observer noted, when a world event occurs, the ability to see workflows could help identify if analysis strategies adapt or stay the same using a meta-analysis technique.
 According to our external collaborators, the current training program relies on a collection of ``good'' analyst suggestions. 
 This definition of good is also debated because it is based on semi-reliable metrics, such as how often an analyst's work is included in another analyst's report.
 However, this does not necessarily indicate quality.
 Visualizing behaviors from a meta-analysis perspective could better inform training programs for analysts instead of relying on inductive reasoning and semi-meaningful metrics.
Ultimately, because few tools support meta-analysis of provenance interactions, the feedback indicated the reviewers found the tool to be thought-provoking and potentially valuable for helping to uncover ecosystem-wide challenges that are symptoms of different internal problems.

To help uncover these ecosystem-wide challenges, we also heard a need for adaptability.
Because there is limited exposure to tools that support meta-analysis of work behavior, we recommend developing tools with flexibility at its core.
This way, users can adapt their existing systems or work with data of various types.
According to the enterprise objectives discussed in the United States National Intelligence Strategy~\cite{natstrat2019}, there is interest in supporting more meta-analysis of work practices across the intelligence community to help identify efficiencies and improve human experiences working with digital tools.
Yet, a few individuals who saw the applied version of our technique mentioned being unfamiliar with the steps involved in completing a meta-analysis.
Some could imagine a context where a user may want to look at what one analyst has done and the selection terms they used. 
In contrast, someone else may care about the whole organization and wants to find the abnormal actions that might be happening. 
Having a pipeline or tool that can dynamically wrangle these different data sources would be helpful.
Additionally, if the ecosystem can consume interaction provenance from various sources, it would likely serve a more diverse set of users.
Therefore, we also suggest that the kinds of data included in a provenance representation or pattern identification method should support a variety of user types, visual representations, and data scenarios.

%% file: sections/08-discussion.tex
\section{Discussion}


\strike{The primary contribution of our research is the presentation of segmentation and summarization methods for generating visual provenance support for communication and collaboration, and the case studies contribute new knowledge about important considerations for how to implement this approach.}
\add{The primary contribution of our research is a literature-based segmentation and summarization technique that reduces the cognitive load required to see the stages of textual analysis sessions.}
\add{Our design choices were made to address the provenance review needs of the intelligence community by being a familiar narrative-like format and can adjust to different audiences and stakeholder types.}
Here, we discuss lessons learned from the case study with the intelligence analysis context, and we consider implications for \strike{broader}\add{similar} applications.
Based on the design iterations and follow-up correspondence with experts following the expert review studies, we  identified a handful of key takeaways and design recommendations for adopting this technique.
While partially grounded in comments provided by experts regarding the interface they worked with, we also make recommendations about provenance representation techniques as a whole.
Our work most generally serves to strengthen the need for more communicative tools in the visualization and data science literature~\cite{Crisan2021Passing}.

\subsection{Takeaway 1: Use Meaningful Segmentation and Hierarchical Segmentation}
Ideally, segmentation and summarization should be inherently interactive\add{\cite{Gadhave_Cutler_Lex_2022,Gratzl2016CLUE}, and when working with different people on an analysis, it's important to establish common ground~\cite{Robinson_2008}}.
Many experts commented on the segmentation approach provided by our tool, requesting additional control over how to segment the data.
They recommended that the boundary between each segment should be ``meaningful,'' not only based on keyword extraction but also on bursts of activity, as an example.
Additional techniques exist for segmentation, including segmentation based on the time between events~\cite{Borkin2013filesystem}, pattern mining based on the order of event types~\cite{Gotz_Wang_Perer_2014}, and Markov modeling for identifying the likelihood of different event transitions~\cite{kwon2021Markov}.
Yet, there is also recent evidence that individuals disagree on the ``best'' segmentation of events~\cite{Sasmita_Swallow_2022}.

As Wang et al.~\cite{Wang2009Temporal} identified, only allowing control of temporal resolution is not sufficient because it does not allow users to select a specific segment, narrow their scope, and learn more about specific periods of the investigation session.
This anecdote demonstrates two key recommendations:
1) ensure the user understands the segmentation method being applied and potentially provide additional ways to chunk investigation sessions, and 
2) when summarizing information for users, these summaries should be hierarchical and allow users to zoom into specific segments for further investigation.


\subsection{Takeaway 2: Pattern Identification for Different Audiences and Purposes}
Our demonstration and expert review sheds light on broader uses about the use of provenance data.
There is a desire to involve more fuzzy nuances to what should be included in the representation (e.g., rapid or repeated searches in relatively short bursts of activity should be collapsed).
In particular, we heard from our experts that in certain use cases, they would want different kinds of data.
For example, team managers need fewer details about the process and more about the general approach.
As provenance data is scaled, there is a need to better support adjusting levels of detail and soft pattern elicitation to more effectively identify areas of interest and support insight~\cite{Gotz_2016}.
Knowing the right patterns to elicit from an interaction log is hard to predict, though.
Suggesting variables of interest or similar visualizations like the voyager tool by Wongsuphasawat et al.~\cite{Wongsuphasawat2016Voyager} could be beneficial because it would guide deeper exploration into the data.
There are also known differences in how people conduct information seeking behaviors given different goals and interfaces~\cite{Edgerly2014Navigational}.

Alternatively, the kinds of information displayed could be user adjustable.
As one expert stated, ``I just want a button to turn off different kinds of information.''
When we added brush-and-link functions to our tool, our experts said it was especially helpful in seeing patterns over time because it provides another way of considering data and its relationships.
\add{In a comparative experiment, Goyal, Leshed, and Fussell~\cite{Goyal_Leshed_Fussell_2013}, show how visualization of uncovered information is more effective for information hand off then relying on written notes.
Choosing an effective design is important for the given context.}
Much like different scientific disciplines may be familiar with certain design conventions~\cite{Gomez2012different}, there are opportunities to adjust a summarization technique to display different information for different audiences.
Having tools that adjust their visualizations for different viewers is a core component of Vi\'egas and Wattenberg~\cite{Viegas2006Communication} concept of Communication-Minded Visualization.
Considering the type of audience and their expected use of the information should influence what kinds of behavior patterns are made available.
Yan et al.~\cite{Yan2021Tessera} describe a technique for automatic pattern identification from interaction logs.
While this technique does not actively consider the type of user reviewing the information and their information needs, they do emphasize how higher-level explanations about why someone did something can be derived from low-level event sequences.
There is also a need for more computer-supported pattern elicitation, including human-in-the-loop approaches that consider the interactions of a user to suggest additional patterns involving similar data~\cite{Zhao2023Chartstory, Han_Schulz_2023guidance}.
This focus on pattern elicitation from lower-level events holds promise for future provenance visualization techniques.


\subsection{Takeaway 3: Similarity of Narrative Visualization and Provenance Visualization} 
Participants generally wanted the prose mode to carry more information than it did, especially more to explain \textit{why} for various segments.
While our implementation opted for a simplified base solution, integrating more advanced NLP techniques could likely better meet user expectations for what would be written on each card. 
Textual blurbs would also benefit from considerations from narrative storytelling. 
Models emphasizing narrative structures like discovery, rescue, or mystery, could likely increase engagement and benefit comprehension~\cite{green2018uniting}.
There is a natural ordering of provenance data, where certain infrequent events carry more semantic information (i.e., search events) than the more common events (i.e., document open events)~\cite{Xu2015Agenda,Bors2019task}.
Since provenance data is essentially hierarchical, it could also benefit from the \textit{minimum description length} principle to aid in summarization~\cite{Veras_Collins_2017, Chen_Xu_Ren_2018}.
There are also applications for considering temporal summary images that include a temporal component with comic strip-style snapshots and annotations for communicating analysis over time~\cite{Bryan_Ma_Woodring_2017TSI}.
Fundamentally, interaction histories carry an analogy to storytelling and narrative techniques because they are essentially episodes over time and could be a valuable research perspective for future work. 

\subsection{Takeaway 4: Adaptability of Provenance Processing}
In addition to the adjustable visualizations for different users, we also recommend the need for flexible data-wrangling tools.
Generating a universal provenance specification for interoperability has been suggested for years~\cite{Bock2020astrographics, Borkin2013filesystem, Groth_Moreau_2013PROV, Gratzl2016CLUE, Gadhave_Cutler_Lex_2022}.
The experts immediately identified many different types of users who could be supported by this and the variety of data sources they would draw from to generate visualizations.
There was an interest in better understanding how to conduct a meta-review. 
Many recognized value in a tool that can dynamically adjust the kind of details included in a report.
Generally, there was a need for pipelines that could flexibly handle interaction logs with various forms.
We demonstrate how relatively small data transformations could be used to allow our technique to interpret five different kinds of interaction logs.
In the wild, legacy systems are commonplace, and building techniques that are agnostic to the variety of data forms yet still interpretable is immediately valuable.

%% file: sections/09-conclusion.tex
\section{conclusion}
We present an algorithmic technique paired with a new representation to better understand user reactions to provenance segmentation and summarization.
Our demonstration with intelligence analysts helped us understand the features of interest related to a segmentation and summarization approach, not to confirm our pipeline was effective.
Using our demonstration as a design artifact, we elicit more realistic user needs for meta-analysis of provenance information.
Our demonstration generates creative interest in the concept of meta-analysis within the intelligence community.
Our experts found the emphasis on time to be less meaningful unless they were to spend more time reviewing the work of their peers. 
Instead, they requested higher levels of data summarization (e.g., topic modeling, action rationale, and final results).
These higher-level summaries are a known challenge to extract from interaction events but represent the goal of analytic provenance et large~\cite{Ram2006Understanding}.
Based on our case study, we call attention to the need for segmentation to have meaningful boundaries and ensure summarization occurs hierarchically so it can be drilled into for additional details.
We call for future work in the directions of additional pattern-matching metric elicitation and interactive filtering to provide brief overviews of work completed for different users.
This inspires future work to better understand how elements like the level of detail, amount of information, and audience types may impact users' understanding of a provenance history. 